\documentclass[nofootinbib,amsfonts,prd,aps,11pt]{revtex4}

\usepackage[T1]{fontenc}

\usepackage{graphicx}
\usepackage{amsmath,amssymb}
\usepackage{hyperref}
\usepackage{graphicx}
\usepackage{subfigure}

\begin{document}


%
%

\title{Evolution in totally constrained models: Schr\"odinger vs. Heisenberg pictures}

\author{Javier Olmedo}

\affiliation{Department of Physics and Astronomy, Louisiana State University, 
202 Nicholson Hall, Tower Dr, Baton Rouge, LA 70803-4001, USA}

%


\begin{abstract}
We study the relation between two evolution pictures that are currently considered for totally constrained theories. Both descriptions are based on Rovelli's evolving constants approach, where one identifies a (possibly local) degree of freedom of the system as an internal time. This method is well understood classically in several situations. The purpose of this paper is to further analyze this approach at the quantum level. Concretely, we will compare the (Schr\"odinger-like) picture where the physical states evolve in time with the (Heisenberg-like) picture in which one defines parametrized observables (or evolving constants of the motion). We will show that in the particular situations considered in this manuscript (the parametrized relativistic particle and a spatially flat homogeneous and isotropic spacetime coupled to a massless scalar field) both descriptions are equivalent. We will finally comment on possible issues and on the genericness of the equivalence between both pictures.
\end{abstract}

\maketitle




\section{Introduction}	

One of the most interesting questions in fundamental physics that is under study nowadays concerns the problem of time. Specifically, in quantum mechanical systems, 
time is often regarded as a mere parameter that rules the evolution, together with the Hamiltonian, by means of the Schr\"odinger equation. However, when the theory in question is generally covariant, like in the case of general relativity, time ceases to be an absolute quantity and the Schr\"odinger equation cannot be applied anymore (at least without 
further considerations). This has been one of the main conceptual problems in quantum gravity since the Hamiltonian of the system is constrained to vanish and the evolution
seems to be frozen within this Schr\"odinger equation. 

Rovelli \cite{rov-evol} showed explicitly that the idea of renouncing to a well defined concept of time at the fundamental level and replacing it by a relational evolution
where the observables of the model evolve with respect to a physical clock variable (instead of a universal time) seems viable also at the quantum level. More explicitly, one identifies one (or several) of the degrees of freedom of the theory with a physical clock and defines the evolution of the system with respect to this physical time. This idea has been studied in many totally constrained theories \cite{mrt,4models,ash-tate}, as well as in more modern approaches for the quantization of general relativity \cite{lqg1,lqg2,obs1,obs2,dust,dust1,dust2} and quantum cosmology \cite{lqc1}. There, one works either in a Heisenberg picture by identifying the observables of the model and represents them on the space of solutions as self-adjoint operators \cite{mrt,4models,rendal}, some of them parametrized in terms of the physical clock. This procedure actually involves in some cases some knowledge about the classical dynamics of the system. Another possibility is to write the constraint equation in an explicit form that resembles the wave equation of either a Klein-Gordon particle \cite{lqc1} or as a Schr\"odinger-like equation \cite{dust,dust1,dust2}. 

In this manuscript we try to reconcile these two pictures in two simple examples: the parametrized relativistic free particle and a spatially flat Friedmann-Robertson-Walker (FRW) spacetime coupled to a massless scalar field. In order to avoid any reference to the previous quantization schemes, we adopt the group averaging technique \cite{raq}. It provides the solutions to the constraint as well as a suitable inner product, where the physical Hilbert space can be constructed and the physical observables of the model represented. We argue how the relational evolution proposed by Rovelli can be implemented in the quantum theory. The physical states become parametrized with respect to a physical time after a measurement of the corresponding physical quantity. We provide a unitary evolution operator with respect to the original physical inner product and the observables of the model. We then construct the corresponding Heisenberg picture, which agrees with the one provided in Ref. \cite{4models} for the relativistic free particle. In the case of the FRW model, we construct the Heisenberg picture that, to the knowledge of the author, has not been explicitly provided before. 

This paper is organized as follows. In Sec. \ref{sec:relativ} we study the parametrized relativistic free particle. We provide a full quantization and the evolution pictures for two choices of time. In Sec. \ref{sec:LQC} we consider a spatially flat FRW spacetime coupled to a massless scalar field. We quantize the model following loop quantum cosmology (LQC) techniques \cite{lqc1}. Finally, we study the two evolution pictures. In Sec. \ref{sec:conc} we give a summary of the results and we comment on different aspects and extensions. 

\section{The parametrized relativistic particle}\label{sec:relativ}

Let us consider a simple totally constrained model corresponding to the parametrized relativistic particle. It is described by the action
\begin{equation} 
S=\int d\tau [p_0 {\dot x}^0+ p{\dot x}  -  N(p_0+ \sqrt{p^2+m^2}) ],
\end{equation}
with $\{x^0,p_0\}=1=\{x,p\}$. Here $N$ is a homogeneous lapse function and $m$ the mass of the particle. One can easily see that a complete set of classical observables is given by $p$ and $q:=x-x^0p/{\sqrt{p^2+m^2}}$ (the momentum and position at initial time). They commute with the classical constraint 
\begin{equation} 
C=p_0+ \sqrt{p^2+m^2},
\end{equation}
and are canonically conjugate, i.e. $\{q,p\}=1$. 

At the classical level, the dynamics of the model can be defined by means of the evolving constants of the motion. If we choose as internal time $T=x^0$, then 
\begin{equation}\label{eq:X-T}
X^0(T)=T, \quad X(T)= q+ {pT}/\sqrt{p^2+m^2},
\end{equation}
where $X^0(T)$ and $X(T)$ are the time and position of the particle, respectively, measured by an observer with proper time $T$. However, if consider an alternative choice of proper time $\tilde T=x^0-ax$ (for the observer),\footnote{In a realistic situation, this would correspond in a very good approximation to a Lorentz transformation with velocity $a=v\ll1$ in units $c=1$.} with $a$ a real constant, one can easily see that the evolving constants of the motion in this case take the form
\begin{equation}\label{eq:X-tilT}
X^0(\tilde T)=\tilde T+aX(\tilde T)={{a q+ \tilde T}\over{1-{{ap}\over \sqrt{p^2+m^2}}}}, \quad X(\tilde T)= {q+ { p\tilde T \over \sqrt{p^2+m^2}}\over{1-{{ap}\over \sqrt{p^2+m^2}}}}.
\end{equation}
In this case, $X^0(\tilde T)$ and $X(\tilde T)$ represent again the time and position of the particle, respectively, but now measured by an observer with proper time $\tilde T$.

\subsection{Quantum mechanics}

We will study the quantization of the model, but let us before summarize the main aspects considered in Ref. \cite{4models}. The basic observables of the model are 
$q$ and $p$, and form a conjugate pair. Then, one adopts a representation of these basic variables as quantum operators on the Hilbert space $L^2[\mathbb{R},dp]$, admitting a representation of the operator algebra $[\hat q, \hat p]=i\hbar$ (in the following we will set $\hbar=1$). Then, in order to represent the evolving constants of the motion, one promotes the observables in Eqs. \ref{eq:X-T} and \ref{eq:X-tilT} to self-adjoint quantum operators on $L^2[\mathbb{R},dp]$. For the choice of time $T$, the observables $\hat X^0(T)$ and $\hat X(T)$ do not involve products of noncommuting operators, so their representation as self-adjoint operators follows straightforwardly. Nevertheless, for the choice of time $\tilde T$, the quantum evolving constants of the motion $\hat X^0(\tilde T)$ and $\hat X(\tilde T)$ involves products of noncommuting operators, so that they must be defined first as symmetric operators on $L^2[\mathbb{R},dp]$ and its self-adjointness must be probed (see Ref. \cite{4models} for a detailed discussion). Completing this last step provides a consistent quantization of the model.

Let us now proceed with the Dirac quantization approach \cite{raq}. We will assume that the kinematical Hilbert space is ${\cal
H}_{kin}= L^2(\mathbb{R}\times\mathbb{R},dp_0dp)$, which is the natural Hilbert space that admits a representation of the operator algebra $[\hat x^0,\hat p_0]=i =[\hat x,\hat p]$. The quantum constraint operator is
\begin{equation} 
\hat C=\hat p_0+ \sqrt{\hat p^2+m^2}.
\end{equation}
The solutions to the constraint, by means of group averaging techniques \cite{raq}, are given by
\begin{equation}\label{eq:aver-sta}
(\Psi|=\int dN \langle e^{iN\hat C}\psi|=\int dp_0dp\delta(p_0+ \sqrt{p^2+m^2})\psi^*(p_0,p)\langle p_0,p|,
\end{equation}
where $| e^{iN\hat C}\psi\rangle:=(2\pi)^{-1}e^{iN\hat C}|\psi\rangle$. The physical inner product is
\begin{equation}\label{eq:inner1}
\langle\Psi|\Phi\rangle_{phy}=(\Psi|\Phi\rangle_{kin}=\int dp_0dp\delta(p_0+ \sqrt{p^2+m^2})\psi^*(p_0,p)\phi(p_0,p).
\end{equation}

Regarding the observables of the model, here denoted by $\hat O$, they can be determined by means of the matrix elements of the corresponding kinematical operators in the kinematical theory. If $\hat o$ is any kinematical operator, then
\begin{equation}\label{eq-obs-kin-phys1}
\langle\Psi|\hat O|\Phi\rangle_{phy}=(\Psi|\left(\hat o|\Phi\rangle_{kin}\right).
\end{equation}

So, for instance, the basic observables of the model can be written (formally) in their respective representations as
\begin{equation}\label{eq-obs-kin-phys2}
\langle\Psi|\hat O|\Phi\rangle_{phy}=\int dp_0dpd\tilde p_0d\tilde p\delta(p_0+ \sqrt{p^2+m^2})\psi^*(p_0,p)\phi(\tilde p_0,\tilde p)\langle p_0,p|\hat o|\tilde p_0,\tilde p\rangle.
\end{equation}
Let us also notice that if we integrate in $p_0$ the right hand side of Eq. \eqref{eq:inner1}, we get
\begin{equation}\label{eq:inner2}
\langle\Psi|\Phi\rangle_{phy}=\int dp\psi^*\Big(p_0(p),p\Big)\phi\Big(p_0(p),p\Big)=\int dp\;\psi^*(p)\phi(p),
\end{equation}
where $p_0(p)=-\sqrt{p^2+m^2}$ and $\phi(p):=\phi\Big(p_0(p),p\Big)$ for any $\phi$ in ${\cal H}_{kin}$.  Then, the physical Hilbert space coincides with $L^2[\mathbb{R},dp]$ in this case. 

This physical picture is, indeed, an abstract theory with no apparent dynamics. In order to achieve a dynamical picture of the model, we need to take into account additional 
considerations. The following sections will be devoted to that. 

\subsubsection{Parametrization for $x^0=T$}

As we saw in the classical theory, $x^0$ can be regarded as a time function. In the quantum theory this corresponds to measuring the observable $\hat X^0$ with a device that is at rest with the particle. The result of the measurement will return the value $T_0$ (a point in the spectrum of $\hat X^0$). After the measurement, the state of the system will be ``projected'' on the corresponding eigenstate\footnote{Interpreting it as a simple projection is not rigorously true as we will see in Sec. \ref{sec:LQC}. In fact, in general, the group averaging can introduce additional weights in the definition of the solutions that must be compensated in the deparametrization process in order to obtain an evolution picture consistent with the traditional Schr\"odinger-like evolution.}, which turns out to be
\begin{equation}\label{eq:state-T0}
\langle\Psi(T_0)|:= \sqrt{2\pi}(\Psi|x^0=T\rangle=\int dp e^{-ip_0(p)T_0}\psi^*(p)\langle p|,
\end{equation} 
where $(\Psi|$ is the averaged state defined in \eqref{eq:aver-sta} once the integral in $p_0$ is carried out. The factor $\sqrt{2\pi}$ has been introduced for convenience. The physical bras \eqref{eq:state-T0}, for any arbitrary function $\psi(p)$ belonging to $L^2[\mathbb{R},dp]$, define the physical kets 
\begin{equation}\label{eq:stateket-T0}
|\Phi(T_0)\rangle=\int dp e^{ip_0(p)T_0}\phi^*(p)|p\rangle,
\end{equation} 
which are all the states belonging to the genuine physical Hilbert space. As we will see below, they will be the starting point in order to construct a Schr\"odinger-like picture for these physical states evolving with respect to the choice of time $T=x^0$. 

Let us start by noticing that, for any two physical states $|\Phi(T_0)\rangle$ and $|\Psi(T_0)\rangle$
\begin{equation}
\langle \Psi(T_0)|\Phi(T_0)\rangle=\int dp \;\psi^*(p)\phi(p)=\langle\Psi|\Phi\rangle_{phy}.
\end{equation} 
So that, the inner product of the parametrized states $|\Phi(T_0)\rangle$ coincides with the physical inner product if it is computed at the same time surface $T=T_0$. We can also define a unitary evolution operator as
\begin{equation}\label{eq:evol-op}
\hat U(T,T_0)=\exp\left\{-i\sqrt{\hat p^2+m^2} (T-T_0) \right\}.
\end{equation}
Here $T$ is any other expectation value of the observable $\hat X^0$ obtained by means of a measurement. One can easily see that this unitary operator maps the original physical state to
\begin{equation}
|\Phi(T)\rangle=\hat U(T,T_0)|\Phi(T_0)\rangle=\int dp e^{ip_0(p)T}\phi(p)|p\rangle.
\end{equation}
The inner product $\langle\Psi(T)|\Phi(T)\rangle$ is independent of $T$ and coincides again with $\langle\Psi|\Phi\rangle_{phy}$. We can define the matrix elements of the physical observables at an arbitrary time $T$ as
\begin{eqnarray}\nonumber
\langle\Psi(T)|\hat P|\Phi(T)\rangle&=&\int dp \;p \psi^*(p) \phi(p), \quad \langle\Psi(T)|\hat P_0|\Phi(T)\rangle=\int dp \; p_0(p) \psi^*(p) \phi(p),\\
\langle\Psi(T)|\hat X|\Phi(T)\rangle&=&\int dp e^{-ip_0(p)T}\psi^*(p) i\partial_p\left(e^{ip_0(p)T}\phi(p)\right).\label{eq:op-mat-elemnt}
\end{eqnarray} 
In order to obtain the last expression, we introduced the resolution of the identity $\hat I=\int dx|x\rangle\langle x|$ in the basis of eigenstates of the operator $
\hat x$, and we assumed as well that $\lim_{p\to \pm\infty}\psi^*(p)\phi(p)=0$.

Regarding the operator $\hat X^0$, since we are working in a Schr\"odinger-like evolution picture with $x^0=T$ as a time, it cannot be defined as an observable in this representation. It must be understood as a parameter of the theory out of which the evolution is defined. 

As it is well known, since the evolution is dictated by a unitary operator  \eqref{eq:evol-op}, it is possible to define a unitarily equivalent Heisenberg picture. In this case, the states are frozen in time, so that any physical state is of the form of 
\begin{equation}
|\Phi\rangle=\int dp\phi(p)|p\rangle,
\end{equation}
while the time dependent operators can be defined by means of Eq. \eqref{eq:op-mat-elemnt} by further developing the derivatives, which yields
\begin{eqnarray}\nonumber
\langle\Psi|\hat P(T)|\Phi\rangle&=&\int dp \; p \psi^*(p) \phi(p), \quad \langle\Psi|\hat P_0(T)|\Phi\rangle=\int dp \; p_0(p) \psi^*(p) \phi(p),\\
\langle\Psi|\hat X(T)|\Phi\rangle&=&\int dp \psi^*(p) \left[i\partial_p\left(\phi(p)\right)+\frac{pT}{\sqrt{p^2+m^2}}\phi(p)\right].\label{eq:op-mat-elemnt-heis}
\end{eqnarray} 
In addition, within this Heisenberg-like picture, one can extend the observable algebra in order to incorporate the time-dependent operator $\hat X^0(T)=T \hat I$. It is worth commenting that this set of quantum operators is in complete agreement with the quantum version of the classical observables defined in $\eqref{eq:X-T}$, as well as with the issue of time and deparametrization discussed in Ref. \cite{ash-tate}.

\subsubsection{Parametrization for $x^0=\tilde T+ax$}

These results are also in agreement with the ones in Ref. \cite{4models}, where an additional choice of time was also considered. In particular, it amounts to introduce the time function $\tilde T=x^0-ax$. In this case one establishes a measuring device in a coordinate system that is not at rest with the particle. In this system we measure the corresponding coordinate $\hat{\tilde X}^0$, obtaining for instance $\tilde T_0$. One can easily define the parametrized states of Eq. \eqref{eq:state-T0}, but now with respect to the time $\tilde T_0$, by means of a unitary transformation, such that
\begin{equation}\label{eq:state-tildT0}
|\Phi(\tilde T_0)\rangle:=\hat {\cal U}(\tilde T_0,T_0)|\Phi(T_0)\rangle=\int dpdx [\mu(p)]^{1/2} \phi(p)e^{ip_0(p)(\tilde T_0+ax)}\frac{e^{ipx}}{\sqrt{2\pi}}|x\rangle ,
\end{equation} 
where
\begin{equation}\label{eq:mu}
\mu(p)=1-\frac{ap}{\sqrt{p^2+m^2}}.
\end{equation}
These parametrized states can also be constructed by carrying out the canonical transformation $[(x^0,p_0),(x,p)]\to[(\tilde x^0,p_0),(x,\tilde p)]$, where $\tilde x^0=x^0-ax$ and $\tilde p = p+ap_0$, at the quantum level in the solutions \eqref{eq:aver-sta}, carry out the measurement of the observable $\hat{\tilde{X}}^0$, and then undo the canonical transformation in order to express the system in terms of the original phase space variables $[(x^0,p_0),(x,p)]$. 

It is possible to see that the inner product of two states at the same initial time $\tilde T_0$ yields, after some straightforward calculations,
\begin{equation}\label{eq:inner-prod-heis-tildT}
\langle \Psi(\tilde T_0)|\Phi(\tilde T_0)\rangle=\int dp \;\psi^*(p)\phi(p)=\langle\Psi|\Phi\rangle_{phy}.
\end{equation} 
So that, the parametrization for the states given in Eq. \eqref{eq:state-tildT0} yields the correct inner product. Since this is valid for any two physical states, this proves in fact that $\hat {\cal U}(\tilde T_0,T_0)$ is a unitary operator. 

Again, the evolution operator for the time $\tilde T$ can also be defined as
 \begin{equation}\label{eq:evol-op-tildT}
 \hat{\tilde  U}(\tilde T,\tilde T_0)=\exp\left\{-i\sqrt{\hat p^2+m^2} (\tilde T-\tilde T_0) \right\},
 \end{equation}
and the evolved states will be
 \begin{equation}
 |\Phi(\tilde T)\rangle=\hat{\tilde  U}(\tilde T,\tilde T_0)|\Phi(\tilde T_0)\rangle=\int dpdx [\mu(p)]^{1/2} \phi(p)e^{ip_0(p)(\tilde T+ax)}\frac{e^{ipx}}{\sqrt{2\pi}}|x\rangle .
 \end{equation}
 
One can check also here that the inner product $\langle\Psi(\tilde T)|\Phi(\tilde T)\rangle$ does not depend on $\tilde T$ and it is equal to $\langle\Psi|\Phi\rangle_{phy}$. 

Then, we can compute the matrix elements of the basic operators of the model. They are given by
\begin{eqnarray}\nonumber
\langle\Psi(T)|\hat P|\Phi(T)\rangle&=&\int dp \;p \psi^*(p) \phi(p), \quad \langle\Psi(T)|\hat P_0|\Phi(T)\rangle=\int dp \; p_0(p) \psi^*(p) \phi(p),\\
\langle\Psi(T)|\hat X|\Phi(T)\rangle&=&\int dp [\mu(p)]^{-1/2}e^{-ip_0(p)\tilde T}\psi^*(p) i\partial_p\left([\mu(p)]^{-1/2}e^{ip_0(p)\tilde T}\phi(p)\right).\label{eq:op-mat-elemnt-tilT}
\end{eqnarray}

As we did above, we will define the Heisenberg picture in this case by means of the unitary operator \eqref{eq:evol-op-tildT}. The corresponding states are time-independent and take the form
 \begin{equation}
 |\Phi\rangle=\int dpdx [\mu(p)]^{1/2} \phi(p)e^{ia p_0(p)x}\frac{e^{ipx}}{\sqrt{2\pi}}|x\rangle.
 \end{equation}
The corresponding inner product agrees with the one in \eqref{eq:inner-prod-heis-tildT}. The matrix elements of the time-dependent operators are 
\begin{eqnarray}\nonumber
 \langle\Psi|\hat P(\tilde T)|\Phi\rangle&=&\int dp \; p \psi^*(p) \phi(p), \quad \langle\Psi|\hat P_0(\tilde T)|\Phi\rangle=\int dp \; p_0(p) \psi^*(p) \phi(p),\\
 \langle\Psi|\hat X(T)|\Phi\rangle&=&\!\!\int dp \psi^*(p) \!\!\left[[\mu(p)]^{-1/2}i\partial_p\left( [\mu(p)]^{-1/2}\phi(p)\right)+\frac{\frac{ p\tilde T}{\sqrt{p^2+m^2}}}{\mu(p)}\phi(p)\right].\label{eq:op-mat-elemnt-heis-tilT}
\end{eqnarray} 
Moreover, we can also extend here the observable algebra in order to incorporate the time-dependent operator $\hat X^0(\tilde T)=\tilde T \hat I+a\hat X(\tilde T)$. Its matrix elements are
 \begin{eqnarray}
\langle\Psi|\hat X^0(\tilde T)|\Phi\rangle&=&\int dp \psi^*(p) \left[a [\mu(p)]^{-1/2}i\partial_p\left( [\mu(p)]^{-1/2}\phi(p)\right)+\frac{\tilde T}{\mu(p)}\phi(p)\right].\label{eq:x0-mat-elemnt-heis-tilT}
 \end{eqnarray} 
Recalling the definition of $\mu(p)$ in Eq. \eqref{eq:mu}, it is easy to realize that the operators \eqref{eq:op-mat-elemnt-heis-tilT} and \eqref{eq:x0-mat-elemnt-heis-tilT} are the quantum analogs of \eqref{eq:X-tilT} on the Hilbert space $L^2[\mathbb{R},dp]$. Besides, this definition is in agreement with the quantum evolving constants provided in Ref.~\cite{4models}. 

Let us comment that the coincidence of the inner product \eqref{eq:inner-prod-heis-tildT} together with these time-dependent operators \eqref{eq:op-mat-elemnt-heis-tilT} and \eqref{eq:x0-mat-elemnt-heis-tilT} with the construction given in Ref. \cite{4models} guarantees their self-adjointness. Let us notice that the only condition that we have imposed in order to properly define them (and which is behind of their self-adjointness) is the fact that $\displaystyle{\lim_{p\to \pm\infty}[\mu(p)]^{-1}\psi^*(p)\phi(p)=0}$. So, the correct choice of the operators is automatically implemented in this quantization. This fact can be traced back to the inner product provided by the group averaging technique. It requires to have a well defined, self-adjoint operator corresponding to the Hamiltonian constraint. So that, the kinematical self-adjoint operators are promoted to self-adjoint Dirac observables (those quantum operators that commute with the self-adjoint quantum constraint).

It is interesting to notice that, as it was discussed by Kuchar in Ref. \cite{kuchar} for the parametrized relativistic free particle, different choices of time function changes the initial Cauchy surface and there is no guaranty that the two quantum theories will be unitarily equivalent. However, the situation changes if one restricts the study to stationary foliations. These are in fact the kind of time functions adopted in our manuscript.

\section{Friedmann-Robertson-Walker spacetime in loop quantum cosmology}\label{sec:LQC}

We will consider now another totally constrained model corresponding to a spatially flat homogeneous and isotropic spacetime coupled to a massless scalar field. In this case, the coordinates in the phase space are given by the scalar field $\phi$  and its canonically conjugated variable $\pi_\phi$, so that $\{\phi,\pi_\phi\}=1$. The geometry, since we will adopt a loop quantization of the model \cite{lqc1,aps,mmo}, will be codified in the physical volume $v=V_0a^3$, with $V_0$ a given fiducial volume and $a$ the scale factor, and the conjugated variable $b$, proportional to the Hubble parameter. Their Poisson algebra is $\{b,v\}=4\pi G\gamma$, where $G$ is the Newton constant and $\gamma$ the Immirzi parameter \cite{lqg1,lqg2}. 

In this case, the classical action of he model is given by
\begin{equation}
S=\int dt \left[\dot\phi\pi_\phi+\frac{1}{4\pi G\gamma}\dot b v -\frac{N}{16\pi G}\left(-\frac{6}{\gamma^2}vb^2+8\pi G \frac{\pi_\phi^2}{v}\right)\right]
\end{equation}
Here, $N$ is a homogeneous lapse function. There are two phase space functions that commute with the constraint (and consequently with the Hamiltonian): $vb$ and $\pi_\phi$. On shell, only one of them is independent. We will not show here the details, but using the equations of motion and the choice of time $\phi=T$, it is possible to construct the basic observables of the model, given by $\Omega_{\pm}=\pm vb$, $\pi_\phi=\frac{\Omega_\pm}{\alpha}$, where $\alpha = \sqrt{\frac{3}{4\pi G\gamma^2}}$, and
\begin{equation}
v^\pm_0=ve^{\mp\sqrt{12\pi G}\phi}.
\end{equation}

As in the relativistic particle, $v^\pm_0$ is the initial volume and it is a constant of the motion. It does not commute with $\Omega_\pm$. In fact, the only nonvanishing Poisson brackets are  $\{\Omega_\pm,v_0^{\pm}\}=4\pi G \gamma v_0^{\pm}$. For instance, we can construct the parametrized observables
\begin{equation}\label{eq:FRW-GR-obs}
V(T) = v_0^\pm e^{\pm\sqrt{12\pi G}T}, \quad B(T)=\frac{\Omega_\pm}{V(T)}=\frac{\Omega_\pm}{v_0^\pm}e^{\mp\sqrt{12\pi G}T}.
\end{equation}

It would be possible to study this spacetime at the quantum level following the ideas of the parametrized relativistic particle studied above. This would correspond to the so-called Wheeler-deWitt quantization. In Ref. \cite{aps} one can find further details, together with a full loop quantization (see also Refs. \cite{lqc1,mmo}). In both cases, a Schr\"odinger-like representation is adopted. We shall notice that a loop quantization involves a polymerization of one (or even several) of the variables. Hence, at the end of the day, we are not quantizing the classical theory we started with. For instance, if we quantize the model from the point of view of a reduced phase space standard quantization, we may look for a representation of the basic observable algebra $[\hat \Omega_\pm,\hat v_0^{\pm}]=4i\pi G \gamma \hat v_0^{\pm}$. Nevertheless, in a loop representation, this basic algebra will be different since the quantum representative of $\hat \Omega_\pm$ would be the polymerized operator corresponding to the classical phase space function $(vb)$.

\subsection{Quantization for periodic functions of $b$}

Let us study the quantization of this model. However, we will not adopt here a genuine polymeric quantization with quasi-periodic functions of the variable $b$. Instead, we will assume periodicity of $b$. This choice will allow us to work within a simpler representation, explicitly solvable and incorporating most of the physical aspects of the original one.

Within this quantization, we start with the operator algebra $[\hat \phi,\hat \pi_\phi]=i$ and $[\widehat{e^{ivb}},\hat v]=4i\pi G\gamma \widehat{e^{ivb}}$. A natural kinematical Hilbert space to represent these operators is ${\cal
H}_{kin}= L^2(\mathbb{R},d\phi)\times L^2([-\pi/\lambda,\pi/\lambda],db)$, where $\lambda$ is a constant proportional to the Planck length (it vanishes for $\hbar\to 0$). In this case, we must specify the periodicity condition in $L^2([-\pi/\lambda,\pi/\lambda],db)$. Given any function $\psi(b)\in L^2([-\pi/\lambda,\pi/\lambda],db)$, we will require $\psi(b=-\pi/\lambda)=e^{-i2\pi\varepsilon}\psi(b=\pi/\lambda)$, with $\varepsilon\in[0,1]$. This involves that the spectrum of $\hat v$ will be $v_n(\varepsilon)=\lambda(\varepsilon+n)$, for $n\in\mathbb{Z}$. So, qualitatively speaking, the label $\varepsilon$ plays a similar role than the one characterizing the superselection sectors in loop quantum cosmology \cite{aps}.

The scalar constraint, after a suitable scaling with $v$,\footnote{Scaling a constraint is subtle since it has important consequences in the quantum theory (see Refs. \cite{klp,louko}). A similar scaling was rigorously implemented in Ref. \cite{mmo} by means of a bijection on the solution space.} can be represented as 
\begin{equation}
\hat C=-\frac{6}{\gamma^2}\hat \Omega^2+8\pi G\hat \pi_\phi^2=24 \pi^2 G^2\left[(-i\partial_b) \frac{\sin(\lambda b)}{\lambda}+\frac{\sin(\lambda b)}{\lambda}(-i\partial_b)\right]^2-8\pi G\partial_\phi^2.
\end{equation}
In addition, in the last equality we give the concrete expression for the constraint in the $(\phi,b)$-representation. It is invariant under $[-\pi/\lambda,0]\to[0,\pi/\lambda]$. So we can restrict the study, for instance, to $b\in [0,\pi/\lambda]$ in the following. This constraint can be diagonalized in the basis of states $|e_k\rangle$ and $|e_\omega\rangle$ of the operators $\hat{\pi}_\phi$ and $\hat \Omega$, respectively, where
\begin{equation}\label{eq:basic-eigenstates}
e_k(\phi)= \langle \phi|e_k\rangle=\frac{1}{\sqrt{2\pi}}e^{ik\phi},\quad e_\omega(b)=\langle b|e_\omega\rangle=\sqrt{\frac{\lambda}{8\pi^2G\gamma}}\frac{e^{i \frac{\omega}{4\pi G\gamma} \log\left[\tan\left(\frac{\lambda b}{2}\right)\right]}}{\sqrt{\sin(\lambda b)}},
\end{equation}
where they fulfill the normalization conditions
\begin{equation}
\langle e_{k'}|e_k\rangle=\delta(k-k'),\quad \langle e_{\omega '}|e_\omega\rangle=\delta(\omega'-\omega).
\end{equation}
Indeed, one can show that the operator $\hat \Omega^2$ is essentially self-adjoint. This can be proved by looking at its deficiency index equations $\hat \Omega^2|\psi_\pm\rangle=\pm i|\psi_\pm\rangle$. In the $b$-representation, they are of the form
\begin{equation}\label{eq:def-eq}
\psi_\pm(b)=\langle b|\psi_\pm\rangle=\sqrt{\frac{\lambda}{8\pi^2G\gamma}}\frac{e^{\mp \frac{1}{4\pi G\gamma} \log\left[\tan\left(\frac{\lambda b}{2}\right)\right]}}{\sqrt{\sin(\lambda b)}}.
\end{equation}
One can easily check that they are not normalizable on $L^2([0,\pi/\lambda],db)$. In addition, the eigenfunctions of $\hat \Omega^2$ are the same than the ones of $\hat \Omega$, and they are normalizable in the generalized sense. We conclude that $\hat \Omega^2$ is essentially self-adjoint.

It is worth commenting that the operator $\hat \Omega^2$  plays the same role than the one in Ref. \cite{mmo} and also the corresponding operator $\hat \Theta$ of Refs. \cite{aps,slqc}. Nevertheless, their difference is a factor ordering and a global constant factor. Indeed, the operator $\hat \Theta$ of Ref. \cite{slqc} coincides with $\hat \Omega^2$ (up to a constant factor) if one adopts a kinematical inner product with continuous measure  $dx=db/\sin(\lambda b)$ instead if $db$. 

The solutions to the constraint can be easily computed and take the form
\begin{equation}\label{eq:aver-sta-lqc}
(\Psi|=\int dkd\omega\delta\left(-\frac{6}{\gamma^2}\omega^2+ 8\pi G k^2\right)\psi^*(\omega,k)\langle e_k,e_{\omega}|.
\end{equation}
The physical inner product is then 
\begin{equation}\label{eq:lqc-inner1}
(\Psi|\tilde\Psi\rangle_{kin}=\int dkd\omega\delta\left(-\frac{6}{\gamma^2}\omega^2+ 8\pi G k^2\right)\psi^*(\omega,k)\tilde\psi(\omega,k).
\end{equation}
As we saw above, in particular Eqs. \eqref{eq-obs-kin-phys1} and \eqref{eq-obs-kin-phys2}, the physical observables of the model can be defined by means of the projection of the kinematical ones on the physical Hilbert space. More explicitly, if we integrate \eqref{eq:lqc-inner1} with respect to $k$, we get
\begin{equation}\label{eq:lqc-inner2}
	(\Psi|\tilde \Psi\rangle_{kin}=\int d\omega\mu(\omega)\left[\psi_+^*(\omega)\tilde\psi_+(\omega)+\psi_-^*(\omega)\tilde \psi_-(\omega)\right],
\end{equation}
where we have defined
\begin{equation}
\mu(\omega)=\frac{\gamma}{\sqrt{12\pi G}|\omega|}.
\end{equation}
Besides, for any $\psi(k,\omega)$ and $\tilde\psi(k,\omega)$ in the kinematical Hilbert space, $\psi_\pm(\omega)=\psi(\omega,k_\pm)$, such that $k_\pm = \pm\alpha|\omega|$ with the constant $\alpha$ defined above. The physical Hilbert space is separated in positive and negative frequencies, where each sector coincides with $L^2[\mathbb{R},|\omega|^{-1}d\omega]$. They are orthogonal and codify the same physical information. This form of the inner product agrees with the one given in Ref. \cite{mmo} and with a particular example of Ref. \cite{embacher}, analogous to the one under study (the Klein-Gordon particle). Nevertheless, its explicit form does not coincide with the one given in Ref. \cite{slqc}. But let us notice that, as explained in Ref. \cite{landsman}, different inner products can be consistently used. There is no tension at all since we can recover the Klein-Gordon inner product of  Ref. \cite{slqc} just by redefining the spectral profiles as $\bar\psi(\omega)=\mu(\omega)\psi(\omega)$.

In the following, we will restrict the study to the positive frequency sector. So, we will omit the labels $(\pm)$. 

The observables can be constructed by means of their definition on the kinematical Hilbert space and the subsequent projection on the physical one. However, we will give more explicit expressions for them below within the two evolution pictures considered in this manuscript. 

The dynamics of this cosmological spacetime has been studied in the context of LQC and the Wheeler-DeWitt quantization \cite{aps,slqc,mmo,klp} and recently reexamined in the formalism of consistent probabilities in Refs. \cite{cprob1,cprob2}. There the constraint equation is interpreted as a second order evolution equation where the scalar field plays the role of time. In consequence it is possible to construct a Schr\"odinger-like picture since the second order equation can be written as a first order one (with positive or negative frequency). The solutions to the constraint can be equipped with a suitable inner product on a given initial slice defined by $\phi=T_0$ together with physical observables and a unitary evolution operator. We will provide an analogous description, but following the ideas presented in the previous section for the relativistic particle.

\subsubsection{Scalar field as a time $\phi=T$}

Since we will consider the scalar field as a time, it is convenient to write the positive frequency solutions to the constraint as
\begin{equation}\label{eq:aver-sta-lqc-pos}
(\Psi|=\int d\omega\mu(\omega)\psi^*(\omega)\langle e_{k=\alpha|\omega|},e_{\omega}|.
\end{equation}

Let us consider a comoving observer that measures the scalar field $\hat \Phi$ with a device. The result of the measurement will return the value $\phi=T_0$ (a point in the spectrum of $\hat \Phi$). After the measurement the state of the system will be 
\begin{equation}\label{eq:state-T0-lqc}
\langle\Psi(T_0)|:=\int d\omega [\mu(\omega)]^{1/2} e^{-i\alpha|\omega| T_0}\psi^*(\omega)\langle e_\omega|.
\end{equation} 
We may notice that this definition is not only a projection of the solution state on the state $|\phi=T_0\rangle$, but also we compensate a factor $[\mu(\omega)]^{1/2}$.\footnote{In the relativistic free particle the density weight in the inner product was already the unit. There, a simple projection of the solutions is enough.} This consideration allows us to get the correct inner product
\begin{equation}\label{eq:inner-T0-lqc}
\langle\Psi(T_0)|\tilde\Psi(T_0)\rangle:=\int d\omega \mu(\omega)\psi^*(\omega)\tilde\psi(\omega).
\end{equation} 

The evolution of the system is determined by a unitary operator whose generator is the conjugated variable to the scalar field $\phi$ (the time of the system) once it has been restricted to the solution space, i.e.,
\begin{equation}\label{eq:evol-op-lqc}
\hat U(T,T_0)=\exp\left\{i\alpha \sqrt{\hat\Omega^2} (T-T_0) \right\}.
\end{equation}
The states at a time $\phi=T$ are then given by
\begin{equation}\label{eq:state-T-lqc}
|\Psi(T)\rangle:=\hat U(T,T_0)|\Psi(T_0)\rangle=\int d\omega [\mu(\omega)]^{1/2} e^{i\alpha|\omega| T}\psi(\omega)|e_\omega\rangle.
\end{equation} 

Regarding the observables of the model, in order to work within the $\omega$-representation, it is more convenient to introduce the set of conjugated observables $\hat\Omega$ an $\hat Y$ fulfilling $[\hat\Omega,\hat Y]=i4\pi G\gamma$. Their expectation values in this representation are 
\begin{align}\label{eq:melem-stat}
	\begin{split}
		&\langle\Psi(T)|\hat\Omega|\tilde\Psi(T)\rangle=\int d\omega \mu(\omega) \omega \psi^*(\omega)\tilde \psi(\omega),\\
		&\langle\Psi(T)|\hat Y|\tilde\Psi(T)\rangle=\int  d\omega [\mu(\omega)]^{1/2} e^{-i\alpha|\omega| T}\psi^*(\omega)(-4i\pi G\gamma)\partial_\omega\left([\mu(\omega)]^{1/2} e^{i\alpha|\omega| T}\tilde \psi(\omega)\right).
	\end{split}
\end{align}
Besides, the momentum conjugated to the scalar field is simply $\hat \Pi_\phi:= \alpha \sqrt{\hat \Omega^2}$. For the operators $\hat B$ and $\hat V$, they can be constructed out of $\hat\Omega$ and $\hat Y$ by means of the relations
\begin{equation}\label{eq:BV-YOMEGA}
\hat B = \frac{2}{\lambda}\arctan\left[e^{\hat Y}\right],\quad \hat V = \frac{\lambda}{2}\left(\frac{1}{\cosh\left[\hat Y\right]}\hat \Omega+\hat \Omega\frac{1}{\cosh\left[\hat Y\right]}\right).
\end{equation}
These set of observables requires that 
\begin{equation}
\lim_{\omega\to\pm\infty}\left\{[\mu(\omega)]^{1/2}\psi^*(\omega)\right\}^{(n)}\left\{[\mu(\omega)]^{1/2}\tilde\psi(\omega)\right\}^{(m)}=0,\quad \forall \;n,m\in\mathbb{N}\cup\{0\}.
\end{equation}
The superscripts mean derivation of order $n$ and $m$, respectively, with respect to $\omega$. Additionally, one can consider to change to the $b$-representation 
\begin{equation}\label{eq:state-b-T-lqc}
|\Psi(T)\rangle=\int db \;\psi (b;T)|b\rangle,
\end{equation} 
where
\begin{equation}\label{eq:wf-b-T-lqc}
\psi (b;T):=\int d\omega [\mu(\omega)]^{1/2} e^{i\alpha|\omega| T}\psi(\omega)e_\omega(b),
\end{equation} 
is simply the Fourier transform of the $\psi(\omega;T):=[\mu(\omega)]^{1/2} e^{i\alpha|\omega| T}\psi(\omega)$ with respect to $e_\omega(b)$ defined in Eq. \eqref{eq:basic-eigenstates}. The expectation values of the operators are then
\begin{align}\label{eq:bvelem-stat}
	\begin{split}
		&\langle\Psi(T)|\hat B|\tilde\Psi(T)\rangle=\int db\;b \psi^*(b;T) \tilde\psi(b;T),\\
		&\langle\Psi(T)|\hat V|\tilde\Psi(T)\rangle=\int db\;\psi^*(b;T)(-4i\pi G\gamma)\partial_b\tilde\psi(b;T).
	\end{split}
\end{align} 

Within the Heisenberg picture, and within the $\omega$-representation, 
the states take the form 
\begin{equation}\label{eq:state-lqc}
|\Psi\rangle:=\int d\omega [\mu(\omega)]^{1/2} \psi(\omega)|e_\omega\rangle.
\end{equation} 
It is straightforward to proof that they yield the inner product \eqref{eq:inner-T0-lqc}. In order to obtain the explicit form of the time-dependent operators $\hat\Omega(T)$ and $\hat Y(T)$, it is enough to take Eq. \eqref{eq:melem-stat} and carry out explicitly some of the derivatives with respect to $\omega$. The result is
\begin{align}\label{eq:melem-stat-heis}
	\begin{split}
		&\langle\Psi|\hat\Omega(T)|\tilde\Psi\rangle=\int d\omega \mu(\omega) \omega \psi^*(\omega)\tilde \psi(\omega),\\
		&\langle\Psi|\hat Y(T)|\tilde\Psi\rangle=\int  d\omega [\mu(\omega)]^{1/2} \psi^*(\omega)(-4i\pi G\gamma)\partial_\omega\left([\mu(\omega)]^{1/2} \tilde \psi(\omega)\right)\\
		&+i\sqrt{12\pi G\gamma^2} T {\rm sgn}(\omega)\mu(\omega)\psi^*(\omega)\tilde\psi(\omega).
	\end{split}
\end{align}

Within this picture, we can also define the time dependent operator corresponding to the momentum conjugated to scalar field as $\hat \Pi_\phi(T):= \alpha \sqrt{\hat \Omega^2(T)}$ and the scalar field itself as $\hat{\Phi} (T)=T\hat I$.

Eventually, the time dependent operators associated with $\hat B$ and $\hat V$ are
\begin{equation}\label{eq:BV-YOMEGA-heis}
\hat B(T) = \frac{2}{\lambda}\arctan\left[e^{\hat Y(T)}\right],\quad \hat V(T) = \frac{\lambda}{2}\left(\frac{1}{\cosh\left[\hat Y(T)\right]}\hat \Omega(T)+\hat \Omega(T)\frac{1}{\cosh\left[\hat Y(T)\right]}\right).
\end{equation}

Therefore, we obtain the corresponding quantum evolving constants of the motion for a spatially flat FRW spacetime coupled to a massless scalar field where the latter plays the role of internal time. Unfortunately, the operators $\hat \Omega$ and $\hat Y$, as well as $\hat B$ and $\hat V$, are formal and they have not been proved to be (essentially) self-adjoint. It will be a matter of future research. 

\section{Conclusions and final remarks}\label{sec:conc}

In this manuscript we study the quantization of two simple mechanical models: the parametrized relativistic free particle and a spatially flat FRW spacetime coupled to a massless scalar field. We construct the quantum operator corresponding to the Hamiltonian constraint and apply the group averaging technique. The physical Hilbert space and the observables of the model are provided. We then construct an evolution picture with respect to a physical time. The unitary evolution is generated by the conjugate variable (restricted to the constraint surface) of the physical time at the quantum level. It is remarkable that experiments in quantum mechanics suggest that time and energy (the generator of time translations) are subject to the Heisenberg uncertainly principle, and our proposal can give a natural explanation, at least in these simple models. We identify the basic quantum observables of the model within a Schr\"odinger-like picture, for each model, and we construct the Heisenberg-like picture thanks to the existence of a well defined unitary evolution operator. 

There are several aspects we would like to comment on the comparison, extension and applications of these results. It is worth commenting that there have been a considerable number of publications regarding the quantization of this FRW model \cite{aps,slqc, mmo,klp,cprob1,cprob2}. Concretely, our analysis has several parallels with the ideas of Ref. \cite{klp}. There, the group averaging technique applied to quantum cosmological spacetimes is compared with a Schr\"odinger picture in the context of loop quantum cosmology. A description in terms of parametrized observables is not explicitly provided within a Heisenberg picture, but only for the Schr\"odinger one. Besides, their work is more general in the sense that it can be directly applied in several isotropic cosmological models, like FRW spacetimes with positive or negative cosmological constant, among others. On the other hand, the proposal of the present manuscript, although applied to a particular cosmological scenario (a spatially flat FRW cosmological scenario coupled to a massless scalar field) with a concrete choice of time, it provides general ideas about how other choices of time function (parametrization of the system) can be carried out in the light of the group averaging quantization and its equivalent version in terms of parametrized observables. This is the first time, to the knowledge of the authors, where a quantization of this spacetimes in terms of parametrized observables has been explicitly constructed and interpreted as a Heisenberg-like picture. Besides, the application of these ideas in some recent studies about the quantization of vacuum spherically symmetric spacetimes \cite{spher1,spher2} will provide a better understanding of the model as well as a connection of the quantization based in parametrized observables with the quantization in terms of the group averaging technique and/or the Schr\"odinger picture commonly employed in cosmological scenarios. 

Regarding the application of our analysis to the case in which $b$ is not a periodic function, in a polymeric quantization, one adopts a representation where $b$ is not well defined as a quantum operator, but only its exponentiation on the kinematical Hilbert space, that now turns out to be non-separable. However, this is not an obstacle since the physical Hilbert space has support on separable subspaces: the so-called superselection sectors. The main difficulty is to define an adequate  operator corresponding to the connection. However, we may notice that the connection has been defined as a function of the operator $\hat Y$, which seems well defined in the quantum theory. So we expect that it will be possible to find a suitable representative for $\hat B$ as well. The technical question that we have not studied in detail concerns the properties of the spectrum of the operator $\hat \Omega$. However, it is analogous to the $\hat \Theta_i$ operators studied in Ref. \cite{bianchi0}, where its spectrum was probed. It is continuous and coincides with the real line. 

The second remark regards alternative choices of time for the system. For the FRW space time model we have regarded the relational time as the scalar field. However, in more general situations, like Bianchi I spacetimes \cite{bianchi}, spherically symmetric gravity \cite{spher1,spher2} and Gowdy cosmologies \cite{gowdy}, the connection is admitted as an internal time of the system. This aspect has been investigated in more detail in Ref. \cite{bianchi} (together with the case in which the triad is regarded as an internal time), but its application to a FRW spacetime has not been yet studied. The results provided in this manuscript can be applied to this situation and compared with the traditional treatments where the scalar field is used as a time. We expect that the physical predictions for both choices of time will be unitarily equivalent. However, the conceptual issue that emerges is whether one can carry out a measurement of the extrinsic curvature, that in this case corresponds to the variable $b$, that is not well defined as an operator. In the asymptotic future and past regions it agrees with the Hubble parameter, but in the high quantum regime it is not possible to identify unequivocally the Hubble parameter with the connection. However, in this high quantum regime, the scalar field can be used again as an internal time and the evolution defined consistently (even for general models where the scalar field interacts with a potential), at least in a very good approximation and for the interesting physical states. Here it would be essential to understand the transformation relating two quantum theories parametrized with respect to two different physical clocks.

The third remark concerns on whether the different choices of time, that have been shown to be equivalent in the cases considered in the parametrized relativistic particle, can affect the predictions in more general situations, particularly in theories with local degrees of freedom. At the classical level one would expect that there will be a well defined symplectomorphism defining the evolution as well as different choices of time should not affect the physical predictions. But the situation can potentially change at the quantum level \cite{kuchar}. For instance, in a canonical quantization, the Schr\"odinger-like picture is constructed on an initial Cauchy surface determined by a given time function. If we consider a different time function, the initial Cauchy surface changes, and there is no guaranty that two quantum theories on each Cauchy surface will be unitarily equivalent. Besides, for a given choice of time, it is not clear whether it is possible to construct an operator providing a unitary evolution in the quantum theory. However, all these issues are far beyond the scope of this manuscript, but are fundamental questions that deserve to be discussed and studied in the future.

In addition, we would like to mention that our construction is based on the group averaging technique \cite{raq}, allowing us to construct the solutions to the quantum constraint as well as it provides a suitable inner product. However, this group averaging requires that the constraint be (an essentially self-adjoint or) a self-adjoint operator, something that is not guaranteed in general. For instance, those theories showing structure functions in the constraint algebra, like general relativity, have a consistent algebra at the quantum level only if one chooses a non-symmetric factor ordering \cite{herm}, spoiling the hermiticity of the corresponding constraints. This impedes the use of group averaging. One can think about formally adopting a group averaging for constraints with a complex spectrum \cite{thiem-ga}, but this fact has not been yet studied in depth. However, the absence of self-adjointness of the constraints does not impede the quantization. There are several examples around in the literature, where one of the most studied ones is an $SL(2,\mathbb{R})$ model with one diffeomorphism and two Hamiltonian constraints \cite{mrt,4models}. The solutions to the constraints can be found and the inner product determined imposing reality conditions on the quantum observables of the model. About the implementation of evolution pictures, like the ones presented in this manuscript, only a Heisenberg-like description has been studied in detail \cite{mrt,4models}. But we still lack a Schr\"odinger-like evolution. Consequently, questions about the unitary equivalence of the two pictures as well as the relation of the quantum descriptions for different choices of time have not been fully investigated yet and will be a matter of future research.

Let us conclude by noticing that the parametrized Dirac observables can also be applied in full quantum gravity for the definition of local observables. Hence, there exists the possibility that the two pictures mentioned in this manuscript are also present there.

\section*{Acknowledgments}

The author is very grateful to Rodolfo Gambini for encouraging him to carry out this study and for inspiring discussions. He also thanks Ivan Agull\'o, Florencia Ben\'itez, Miguel Campiglia and Jorge Pullin for discussions, as well as the anonymous Referee for suggestions and improvements of the present manuscript. The author acknowledges the support provided by the grants MICINN/MINECO FIS2011-30145-C03-02 and FIS2014-54800-C2-2-P from Spain, NSF-PHY-1305000 from USA and Pedeciba from Uruguay.

\end{document}